\documentstyle[12pt]{article}
\def\ie{{\em i.e.}}
\def\eg{{\em e.g.}}

\catcode`\@=11 

\def\lsim{\mathrel{\mathpalette\@versim<}}
\def\gsim{\mathrel{\mathpalette\@versim>}}
\def\@versim#1#2{\vcenter{\offinterlineskip
    \ialign{$\m@th#1\hfil##\hfil$\crcr#2\crcr\sim\crcr } }}
\def\etal{{\em et. al.}}
\def\JL{J. L. Lopez}
\def\DVN{D. V. Nanopoulos}

\def\t1{{\tilde 1}}

\def\GeV{\,{\rm GeV}}
\def\TeV{\,{\rm TeV}}

\def\ipb{\,{\rm pb}^{-1}}

\def\NPB#1#2#3{Nucl. Phys. B {\bf#1} (19#2) #3}
\def\PLB#1#2#3{Phys. Lett. B {\bf#1} (19#2) #3}

\def\PRD#1#2#3{Phys. Rev. D {\bf#1} (19#2) #3}
\def\PRL#1#2#3{Phys. Rev. Lett. {\bf#1} (19#2) #3}

\def\hepph#1{{\tt hep-ph/#1}}

\begin{document}
\begin{flushright}
\baselineskip=12pt
CERN-TH/95-314\\
DOE/ER/40717--22\\
CTP-TAMU-46/95\\
ACT-17/95\\
\tt hep-ph/9512288
\end{flushright}

\begin{center}
\vglue 1cm
{\Large\bf Supersymmetry and $\rm R_b$ in the light of LEP 1.5\\}
\vglue 0.75cm
{\Large John Ellis,$^1$ Jorge L. Lopez,$^2$ and D.V. Nanopoulos$^{3,4}$\\}
\vglue 0.5cm
\begin{flushleft}
$^1$CERN Theory Division, 1211 Geneva 23, Switzerland\\
$^2$Department of Physics, Bonner Nuclear Lab, Rice University\\ 6100 Main
Street, Houston, TX 77005, USA\\
$^3$Center for Theoretical Physics, Department of Physics, Texas A\&M
University\\ College Station, TX 77843--4242, USA\\
$^4$Astroparticle Physics Group, Houston Advanced Research Center (HARC)\\
The Mitchell Campus, The Woodlands, TX 77381, USA\\
\end{flushleft}
\end{center}

\vglue 0.5cm
\begin{abstract}
We re-examine the possible magnitude of the supersymmetric
contribution to $R_b\equiv\Gamma(Z^0\to\bar b b)/\Gamma(Z^0\to{\rm all})$
in the light of the constraints imposed by the absence of light charginos at
LEP~1.5, implementing also other available phenomenological constraints. We
find the supersymmetric contribution to be $R^{\rm susy}_b < 0.0017$, and
discuss the extent to which this upper bound could be strengthened by future
constraints on the chargino and top-squark masses. Such values of $R^{\rm
susy}_b$ tend to disfavor a supersymmetry explanation of the apparent $R_b$
discrepancy.
\end{abstract}
\vspace{1cm}
\begin{flushleft}
\baselineskip=12pt
CERN-TH/95-314\\
DOE/ER/40717--22\\
CTP-TAMU-46/95\\
ACT-17/95\\
December 1995
\end{flushleft}

\vfill\eject
\setcounter{page}{1}
\pagestyle{plain}
\baselineskip=14pt

LEP~1 has, unfortunately, provided a showcase for the Standard Model, which has
been tested successfully down to the {\it per mille} level. The measurements
have proved to be sensitive to quantum corrections within the Standard Model,
which have enabled the mass of the top quark to be predicted accurately, and
may now be sensitive to the mass of the Higgs boson \cite{EFL}.
The only possible blots on
the Standard Model's copybook have been suggested by the LEP measurements of
$Z^0$ decays into $\bar b b$ and $\bar c c$. The preliminary measurements of
$R_{b,c}\equiv\Gamma(Z^0\to\bar b b,\bar c c)/\Gamma(Z^0\to{\rm all})$ reported
at the Brussels and Beijing conferences \cite{Rbexp} disagree {\it prima facie}
with the Standard Model at the levels of 3.7 and 2.5 standard deviations,
respectively. Even if $R_c$ is fixed to its Standard Model value, at a
considerable cost in $\chi^2$, the LEP~1 measurement of $R_b$ still disagrees
with the Standard Model at the level of 3 standard deviations. It may well be
that the apparent discrepancy is in fact due to a misestimation of the
uncertainties associated with the simulation of the $\bar b b$ and $\bar c c$
final states, but it has been seductive to speculate that some new physics
beyond the Standard Model may be coming into play.

One such speculation has been supersymmetry \cite{Rbsusy}, and two specific
scenarios to explain the $R_b$ discrepancy (but not the $R_c$ one) have been
proposed. One has invoked a light chargino $\chi^{\pm}_1$ and a light
top-squark $\tilde t_1$ close to the kinematic limits already excluded by new
particle searches at LEP~1 \cite{KKW}, and the other a light pseudoscalar Higgs
boson $A$ \cite{GJS}. These have inspired the hope in some quarters that one or
more of these supersymmetric particles might be produced at LEP~2, and
conceivably already in the intermediate-energy LEP~1.5 run recently completed.
It should be pointed out, though, that it is has proved difficult in specific
models to obtain a supersymmetric contribution to $R_b$ large enough to remove
the apparent discrepancy, once one applies plausible phenomenological or
theoretical constraints \cite{WLN,Others}.

Preliminary results of the first part of the LEP~1.5 run have now been
announced by the four LEP collaborations, and, to paraphrase Sherlock
Holmes, ``the curious incident was that the dog did nothing". Specifically, all
the four LEP collaborations have reported preliminary lower limits on the mass
of the lighter chargino \cite{LEP15}: $m_{\chi^{\pm}_1}\gsim65\GeV$ if
$m_{\chi^{\pm}_1}-m_{\chi^0_1}\gsim10\GeV$ (with some dependence on the
sneutrino mass), where the $\chi^0_1$ is the lightest
neutralino, which is assumed to be the lightest supersymmetric particle (LSP).
Many people are aware that this news is particularly disappointing for
advocates of the light ($\chi^{\pm}_1, \tilde t_1$) interpretation of the $R_b$
anomaly. The purpose of this note is to quantify the upper limit on the
possible supersymmetric contribution to $R_b$ in the light of this preliminary
LEP~1.5 result, as well as recent D0 constraints on the $\tilde t_1$ mass and
updates of other experimental constraints on possible sparticle masses, limits
on possible new physics effects in $Z^0,t$ and $b$ decay, and the absence of
the lightest supersymmetric Higgs boson.

To set the scene for our study, we first recall that the Standard Model
contribution to $R_b$ (for $m_t=175\GeV$) is $R_b^{\rm SM}=0.2157$ \cite{RbSM},
whereas the reported experimental value (with $R_c$ constrained to the Standard
Model value) is $R_b^{\rm exp}=0.2205\pm0.0016$ \cite{Rbexp}. This means that a
value of $R^{\rm susy}_b\ge0.0020$ would bring the supersymmetric $R_b$
prediction within the 95\% C.L. interval, whilst a contribution $R^{\rm
susy}_b\ge0.0030$ would bring the prediction within one sigma of the
experimental value.

In this note we consider the supersymmetric contributions to $R_b$ in the
regime of light chargino and top-squark masses and small values of $\tan\beta$,
where they may be enhanced \cite{KKW}. Enhancements to $R_b^{\rm susy}$ may
also occur for small values of the pseudoscalar Higgs mass ($m_A$)
and large values of $\tan\beta$ \cite{GJS}, but this scenario now appears to be
disfavored \cite{WK}, and we do not consider it in what follows. The dominant
contribution\footnote{We take the mass of the charged
Higgs boson ($H^\pm$) to be large, so as to minimize the contribution to
$R^{\rm susy}_b$ from the $H^\pm-t$ loop, which is always negative.
This means that our results are conservative upper bounds.} to $R^{\rm susy}_b$
then depends on six parameters: those that parametrize the chargino sector
($M_2,\mu,\tan\beta$), the top-squark masses ($m_{\tilde t_1}<m_{\tilde t_2}$),
and their mixing angle ($\theta_{\tilde t}$). We work in the context of the
general Minimal Supersymmetric Standard Model (MSSM), without assuming
{\em a priori} any relationship among these parameters that might result from
unification conditions or dynamical models.

Following Ref.~\cite{WLN}, we first sample a large number of six-plet choices
of parameters, with those parameters that have the dimension of mass allowed to
take random values in the interval $(0\to250)\GeV$, and with $\tan\beta$
restricted to the range $1\to5$. The total sample of approximately 365K
six-plets is restricted in such a way that the most elementary LEP~1 lower
bounds ($m_{\chi^\pm_1},m_{\tilde t_1}>45\GeV$) are satisfied.
We then find a total of 1000 six-plets that yield $R^{\rm susy}_b\ge0.0020$.
To examine in more detail the region of low values of $\tan\beta$,  we have
also generated and studied a ``low-$\tan\beta$" sample (91K six-plets), for
which $\tan\beta$ is restricted to the range $1\to1.5$.

In order to determine the upper bound on $R^{\rm susy}_b$, we apply a series
of experimental constraints to our large six-plet sample, as follows:
\begin{enumerate}
\item The invisible $\Gamma(Z\to\chi^0_1\chi^0_1)$ width should be less
than 3.9~MeV, as can be inferred from the most recent LEP
result $\delta\Gamma_{\rm inv}=(-1.5\pm2.7)$~MeV \cite{Rbexp}.
\item The branching ratio $B(Z\to\chi^0_1\chi^0_2)$ should not exceed
$10^{-4}$ \cite{L3}.
\item The more restrictive LEP~1 lower limit on the chargino mass:
$m_{\chi^\pm_1}>47\GeV$, valid for $m_{\chi^0_1}<42.5\GeV$ and for the
higgsino-like chargino \cite{ALEPH}
required for an enhancement in $R^{\rm susy}_b$.
\item The lightest Higgs boson should be heavier than
the LEP~1 limit ($m_h\gsim40\GeV$). The mass of this Higgs boson
acquires a large quantum correction at the one-loop level, which is
dominated by the top--top-squark loop \cite{ERZ}. Casting the one-loop
correction in terms of
the observable top-squark parameters ($m_{\tilde
t_{1,2}},\theta_{\tilde t}$) alone, one obtains \cite{LNhiggs}
\begin{eqnarray}
(m^2_h)^{\rm max}&=&M^2_Z\Biggl\{\cos^22\beta
+\gamma \left({m_t\over M_Z}\right)^4
\Biggl[\ln{m^2_{\tilde t_1}m^2_{\tilde t_2}\over m^4_t}\\
&&+(m^2_{\tilde t_1}-m^2_{\tilde t_2}){\sin^22\theta_{\tilde t}\over2m^2_t}
\ln{m^2_{\tilde t_1}\over m^2_{\tilde t_2}}
+(m^2_{\tilde t_1}-m^2_{\tilde t_2})^2
\left({\sin^22\theta_{\tilde t}\over4m^2_t}\right)^2
g\left({m^2_{\tilde t_1}\over
m^2_{\tilde t_2}}\right)\Biggr]\Biggr\}\nonumber
\end{eqnarray}
with $\gamma=3\alpha/(4\pi\sin^2\theta_W\cos^2\theta_W)$ and
$g(r)=2-{r+1\over r-1}\ln r$ [$g(1)=0$, $g(r)\le0$].
Other one-loop corrections and the largest of the two-loop corrections
are not expected to be large \cite{HiggsLoops}, and are probably no
larger than uncertainties in the approximations used, so we do not
incorporate them.
\item The branching ratio $B(b\to s\gamma)$ should fall in the range
$(1-4)\times10^{-4}$. This interval is a conservative interpretation of
the latest CLEO result $B(b\to s\gamma)^{\rm exp}=(2.32\pm0.57\pm0.35)
\times10^{-4}$ \cite{CLEO}, which should cover the theoretical
uncertainties in the calculation of $B(b\to s\gamma)$, principally
due to higher-order perturbative QCD corrections in the Standard Model
contribution.
\item The branching ratio $B(t\to bW)$ has been determined by CDF to be
$0.87^{+0.13}_{-0.32}$ \cite{CDF}. We therefore require $B(t\to{\rm
new})<0.45$, where ``new" includes in our case the $t\to\tilde
t_{1,2}\chi^0_{1,2}$ decay channels, when kinematically allowed.
More restrictive upper limits on $B(t\to{\rm new})$ have been considered
elsewhere \cite{WLN,MY}.
\item The D0 Collaboration has included a region in the
$(m_{\chi^0_1},m_{\tilde t_1})$ space, assuming that
$m_{\tilde t_1}<\{m_{\chi^\pm_1},m_{\tilde\ell},m_{\tilde\nu}\}$
\cite{D0}. These restrictions insure that the dominant $\tilde t_1$ decay
mode is via the one-loop process $\tilde t_1\to c\chi^0_1$.
\item The new LEP~1.5 lower limit on the chargino mass $m_{\chi^\pm_1}
\gsim65\GeV$, valid as long as $m_{\chi^\pm_1}-m_{\chi^0_1}\gsim10\GeV$
\cite{LEP15}. A more precise formulation of the limit must await the
publication of their results by the LEP collaborations: it depends on the
sneutrino mass and on the wino/higgsino content of the chargino. It seems to us
that the above limit is conservative, applying when the sneutrino is heavy, or
when the chargino is higgsino-like, which is the case of relevance for
obtaining a large value of $R^{\rm susy}_b$. We also discuss later the effect
of decreasing the restriction on the chargino-neutralino mass difference
to about 5 GeV, as might be achieved in the final analysis.
\end{enumerate}
Motivated by the requirement that any stable supersymmetric relic particle
should be electromagnetically neutral and have no strong interactions
\cite{EHNOS}, we also require that neither the lightest top-squark nor the
lightest chargino should be the lightest supersymmetric particle, \ie,
$\{m_{\chi^\pm_1},m_{\tilde t_1}\}>m_{\chi^0_1}$.

After running our large sample of six-plets through the above set of
experimental and theoretical constraints, we find that no points with
$R^{\rm susy}_b>0.0020$ survive. The main reason for this result is the new
LEP~1.5 constraint on the chargino mass. This could have been anticipated, as
Refs.~\cite{KKW,WLN,WK}, which did not have access to the new data, found
regions of parameter space with $R^{\rm susy}_b>0.0020$, even after enforcing
most of the constraints enumerated above. We conclude that a
supersymmetric solution to the $R_b$ anomaly is less likely in the light of
LEP~1.5. This conclusion holds for both our ``regular" sample and our
``low-$\tan\beta$" sample. Moreover, these results rely only on the
present LEP~1.5 result, with the chargino-neutralino mass difference
required to be more than 10 GeV, and are in fact independent of the constraint
on the Higgs-boson mass (item 4 above). We should add that our full sample
contains a small fraction of points with very low values of the neutralino
masses (few GeV), which manage to pass all LEP~1 constraints (see also
\cite{WLN,Feng}) and are not subjected to the known limits on the gluino
mass as we do not impose the GUT relation among gaugino masses. These points
are, however, all excluded by either the $B(b\to s\gamma)$ constraint (item 5)
or the LEP~1.5 constraint (item 8).

\begin{figure}[p]
\vspace{5in}
\includegraphics{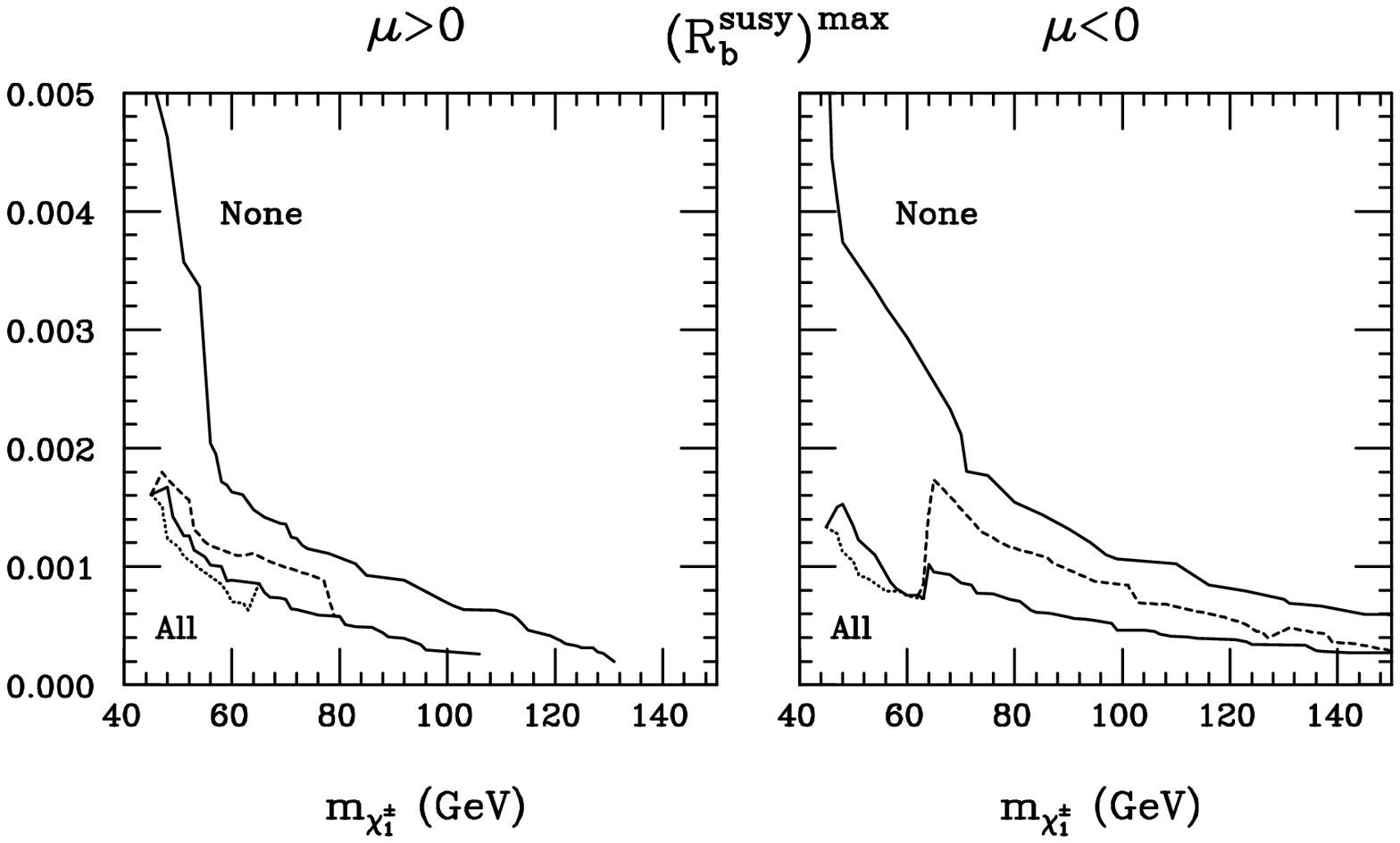}
\caption{The maximum attainable value of $R^{\rm susy}_b$ versus the chargino
mass for both signs of $\mu$, when no constraint has been applied (``None") and
when all the constraints described in the text have been applied (``All"). The
dashed lines indicate the effect of not enforcing the Higgs-mass constraints,
and the dotted lines indicate the possible further restriction should future
LEP~1.5 searches exclude a chargino-neutralino mass down to about 5 GeV.}
\label{fig:Rbmax}
\end{figure}

Next we look for the largest achievable values of $R^{\rm susy}_b$. In
Fig.~\ref{fig:Rbmax}, we show $(R^{\rm susy}_b)^{\rm max}$
as a function of the lightest chargino mass ($m_{\chi^\pm_1}$),
for both signs of $\mu$. The top curves (``None") give the raw results obtained
from the full sample of parameter six-plets, whereas the (solid) bottom
curves (``All") give the limiting values when {\em all} the above constraints
are applied, in which case we find the absolute upper limit
\begin{equation}
R_b^{\rm susy}<0.0017\ .
\label{absolute}
\end{equation}
Of particular importance in excluding values of $\tan\beta\approx1$ is the
Higgs mass constraint (item 4 above). As has already been mentioned, this
constraint is worthy of further theoretical refinement, and may soon be
strengthened by LEP itself. The effect of not enforcing this constraint is
represented by the dashed lines in Fig.~\ref{fig:Rbmax}. Note that this
constraint is superseded by the LEP~1.5 constraint for
$m_{\chi^\pm_1}\lsim65\GeV$. We also display as dotted lines the further
restriction that may be obtained should the LEP~1.5 be strengthened to exclude
chargino-neutralino mass differences down to about 5 GeV, assuming that the
lower bound on the chargino mass remains at 65 GeV.  We note that if it were
possible to obtain an absolute lower bound of 65 GeV on the chargino mass, then
only values of $R^{\rm susy}_b<0.0010$ would be possible. Future runs at LEP~2
energies should be able to probe chargino masses as large as 90 GeV, which
would imply $R^{\rm susy}_b<0.0005$, should no chargino signal be
observed.\footnote{Note that just as LEP~1.5 has not been able to set an
absolute lower limit on the chargino mass because of the experimental
limitation of a minimal chargino-neutralino mass difference, the same could
happen at LEP~2 energies. This limitation may be overcome by resorting to a
hard photon tag, as recently discussed in Ref.~\cite{CDG}.}

The Tevatron should also be able to constrain $(R^{\rm susy}_b)^{\rm max}$
by setting lower limits on the chargino mass. Indeed, D0 has just released its
first limits on chargino-neutralino production and decay into trilepton final
states \cite{D03l}. The limits are on the trilepton rates, \ie, $\sigma(p\bar
p\to\chi^\pm_1\chi^0_2X)\cdot B(\chi^\pm_1\to \ell)\cdot
B(\chi^0_2\to 2\ell)$, which can be translated into limits on the chargino mass
once one calculates the trilepton branching ratio. The latter depends on the
detailed spectrum of sleptons and squarks (which we do not consider), and may
be enhanced if there are light sleptons \cite{LNWZ}, in which case the D0
limits imply $m_{\chi^\pm_1}\gsim55\GeV$ \cite{D03l}. The possibility of light
sleptons will soon be explored at LEP, and the D0 sensitivity to trileptons is
expected to increase significantly once the full data set is analyzed.

\begin{figure}[p]
\vspace{5in}
\includegraphics{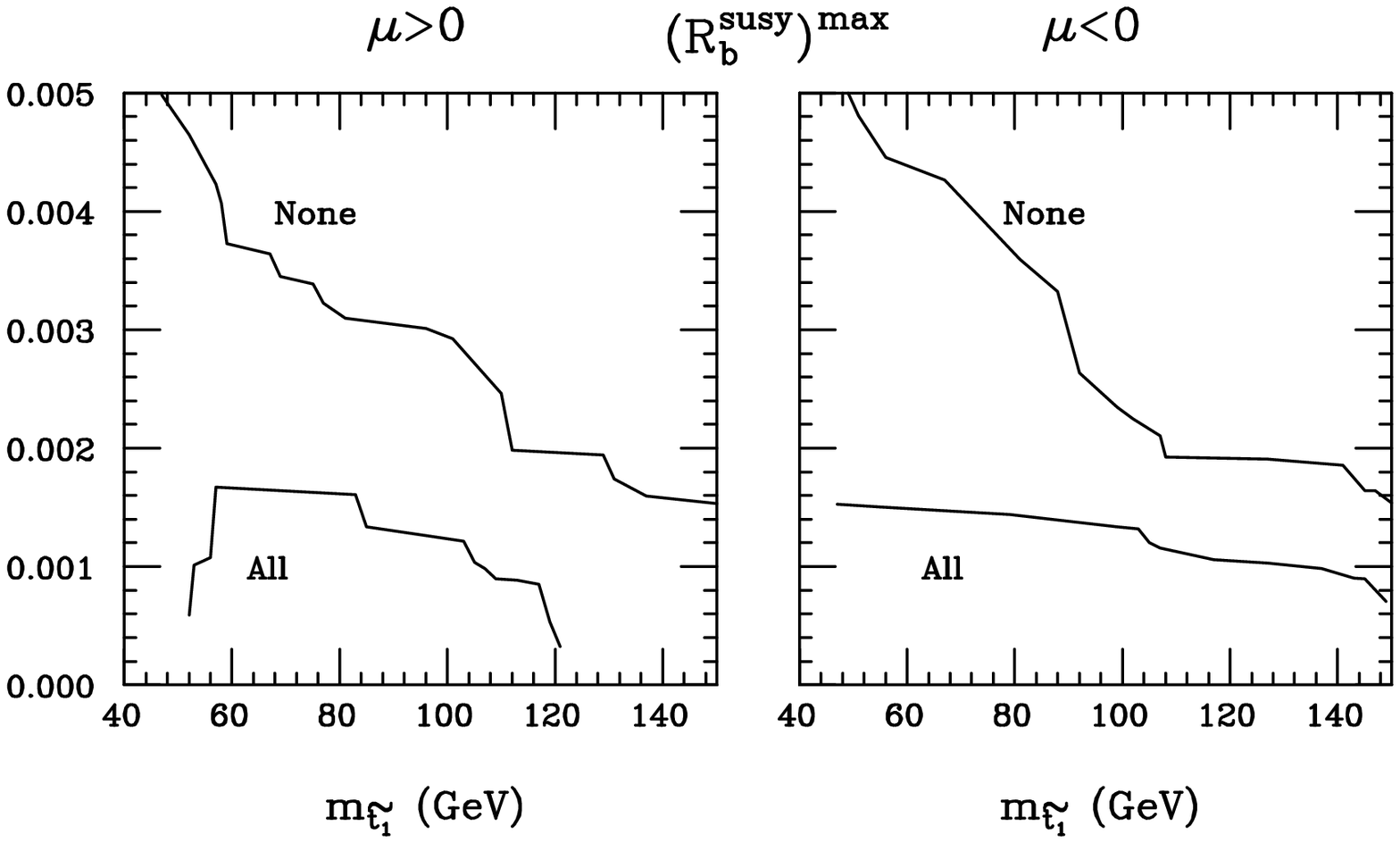}
\caption{The maximum attainable value of $R^{\rm susy}_b$ versus the top-squark
mass for both signs of $\mu$, when no constraint has been applied (``None") and
when all constraints have been applied (``All"). It can be seen from this
plot how the expected future direct limits on $m_{\tilde t_1}$ from the
Tevatron will constrain $R^{\rm susy}_b$ further.}
\label{fig:Rbmax-stop}
\end{figure}

With a view to present and future top-squark searches at LEP and the Tevatron,
we have also studied the dependence of $(R^{\rm susy}_b)^{\rm max}$ on the
lightest top-squark mass. This is shown in Fig.~\ref{fig:Rbmax-stop} for the
``None" and ``All" cases (with the Higgs mass constraint included and allowing
a chargino-neutralino mass difference of up to 10 GeV). Direct top-squark
searches at the Tevatron are underway, but so far have concentrated on
top-squark decays via $t\to c\chi^0_1$. This decay is dominant as long as
$m_{\tilde t_1}<\{m_{\chi^\pm_1}, m_{\tilde\ell},m_{\tilde\nu}\}$. With this
restriction, D0 has excluded a region in the $(m_{\chi^0_1},m_{\tilde t_1})$
plane \cite{D0}. This region is not very constraining for our present purposes,
but it is expected that top-squark masses as large as 130 GeV could be explored
with the data ($\sim100\ipb$) already accumulated. As Fig.~\ref{fig:Rbmax-stop}
shows, a lower bound of this magnitude would impose new severe restrictions on
the allowed values of $R^{\rm susy}_b$.

We have also explored the dependence of $(R^{\rm susy}_b)^{\rm max}$ on
$B(b\to s\gamma)$ and $B(t\to{\rm new})$. We find that more stringent
experimental limits will decrease further the size of the allowed region
in parameter space, but will not necessarily impose important new restrictions
on $(R^{\rm susy}_b)^{\rm max}$.

Requiring rather light top-squark masses may entail a degree of fine-tuning
in the top-squark mass matrix, such as large values of $A_t$. In the limit
$\tan\beta\approx1$ this situation may lead to minima of the electroweak
scalar potential that break electric or color charge \cite{LP}. We do not
include these constraints in the present analysis, as these would only further
constrain the allowed region of parameter space.

Before concluding, we note that imposing further theoretical
constraints on the parameter space, such as those that follow from
universal supersymmetry breaking masses at the GUT scale and radiative
electroweak breaking, tend to reduce $(R^{\rm susy}_b)^{\rm max}$ very
substantially \cite{KKW,WLN}. Consulting Fig.~1 in Ref.~\cite{WLN}, one can see
that $R^{\rm susy}_b\lsim0.0002$, after the new LEP~1.5 limit is imposed.

Even without imposing such additional theoretical constraints, the
central result (\ref{absolute}) of our analysis suggests that the previously
most plausible supersymmetric scenario for accommodating the apparent anomaly
in $R_b$ is now so severely constrained that it no longer appears able
to resolve this experimental discrepancy with the Standard Model. In the
absence of any other promising explanation from beyond the Standard Model,
it may be necessary to review carefully the calculation and simulation of
the Standard Model contributions to $R_b$ and related measurements. LEP 1.5
has done much to clarify the prospects of a supersymmetric resolution of this
LEP 1 anomaly, and further stages of LEP should be able to cement our
conclusion.

\section*{Acknowledgments}
We thank Carlos Wagner and James White for helpful discussions.
The work of J.~L. has been supported in part by
DOE grant DE-FG05-93-ER-40717, and that of
D.V.N. has been supported in part by DOE grant DE-FG05-91-ER-40633.

\clearpage

\clearpage


\newpage
\begin{thebibliography}{99}
\bibitem{EFL}J. Ellis, G.L. Fogli, and E. Lisi, CERN-TH/95-202
(\hepph{9507424}), and references therein.
\bibitem{Rbexp}P.~Renton, Rapporteur talk at the International
Symposium on Lepton and Photon Interactions at High Energies,
High Energy Physics, Beijing (August 1995), Oxford preprint
OUNP-95-20 (1995).
\bibitem{Rbsusy}
G. Altarelli and R. Barbieri, \PLB{253}{90}{161};
M.~Boulware, D.~Finnel, \PRD{44}{91}{2054};
A.~Djouadi, G.~Girardi, C.~Vergzegnassi, W.~Hollik and F.~Renard,
\NPB{349}{91}{48};
G. Altarelli, R. Barbieri, and S. Jadach, \NPB{369}{92}{3};
G. Altarelli, R. Barbieri, and F. Caravaglios, \NPB{405}{93}{3};
G. Altarelli, R. Barbieri, and F. Caravaglios, \PLB{314}{93}{357}.
\bibitem{KKW}J.~D.~Wells, C.~Kolda, and G.~L.~Kane, \PLB{338}{94}{219}.
\bibitem{GJS}D.~Garcia, R.~Jimenez, and J.~Sola, \PLB{347}{95}{321};
D.~Garcia and J.~Sola, \PLB{357}{95}{349}.
\bibitem{WLN}X. Wang, \JL, and \DVN, \PRD{52}{95}{4116}.
\bibitem{Others} G. Kane, R. Stuart, and J. Wells, \PLB{354}{95}{350};
E. Ma and D. Ng, \hepph{9508338};
Y. Yamada, K. Hagiwara, and S. Matsumoto, \hepph{9512227}.
\bibitem{LEP15}L. Rolandi, H. Dijkstra, D. Strickland and
G. Wilson, representing the ALEPH, DELPHI, L3 and OPAL
collaborations, Joint Seminar on the First Results from LEP~1.5,
CERN, Dec.~12th, 1995.
\bibitem{RbSM}
A. Akhundov, D. Bardin, and T. Riemann, \NPB{276}{86}{1};
J. Bernabeu, A. Pich, and A. Santamaria, \PLB{200}{88}{569};
W. Beenaker and W. Hollik, Z. Phys. C40, 141(1988);
F. Boudjema, A. Djouadi, and C. Verzegnassi, \PLB{238}{90}{423};
A. Blondel and C. Verzegnassi, \PLB{311}{93}{346}.
\bibitem{WK}J. Wells and G. Kane, \hepph{9510372}.
\bibitem{L3}M.~Acciarri, \etal\ (L3 Collaboration), \PLB{350}{95}{109}.
\bibitem{ALEPH}See, \eg, D. Decamp, \etal\ (ALEPH Collaboration),
Phys. Reports {\bf216} (1992) 253.
\bibitem{ERZ}Y. Okada, M. Yamaguchi, and T. Yanagida,
Prog. Theor. Phys. {\bf
85} (1991) 1 and \PLB{262}{91}{54}; J. Ellis, G. Ridolfi, and F. Zwirner,
\PLB{257}{91}{83} and \PLB{262}{91}{477}; H. Haber and R. Hempfling,
\PRL{66}{91}{1815}.
\bibitem{LNhiggs}\JL\ and \DVN, \PLB{266}{91}{397}.
\bibitem{HiggsLoops}M. Diaz and H. Haber, \PRD{46}{92}{3086};
R. Hempfling and A. Hoang, \PLB{331}{94}{99};
M. Carena, J. Espinosa, M. Quiros, and C. Wagner,
\PLB{355}{95}{209}; M. Carena, M. Quiros, and C. Wagner, \hepph{9508343}.
\bibitem{CLEO}T.E. Browder and K. Henscheid, University of Hawaii and Ohio
State University preprint, UH 511-816-95 and OHSTPY-HEP-E-95-010 (1995), to
appear in {\it Progress in Nuclear and Particle Physics}, Vol. 35.
\bibitem{CDF} J. Incandela (CDF Collaboration), FERMILAB-CONF-95-237-E
(July 1995).
\bibitem{MY}S. Mrenna and C.-P. Yuan, \hepph{9509424}.
\bibitem{D0} S. Abachi, \etal\ (D0 Collaboration), ``Search for light top
squarks in $p\bar p$ collisions at $\sqrt{s}=1.8\TeV$", December 1995
(submitted to Phys. Rev. Lett.).
\bibitem{EHNOS}J. Ellis, J.S. Hagelin, D.V. Nanopoulos, K.A. Olive
and M. Srednicki, \NPB{238}{84}{453}.
\bibitem{Feng}J. Feng, N. Polonsky, and S. Thomas, \hepph{9511324}.
\bibitem{CDG}C.-H.~Chen, M. Drees, and J. Gunion, \hepph{9512230}.
\bibitem{D03l}S. Abachi, \etal\ (D0 Collaboration), {\tt hep-ex/9512004}.
\bibitem{LNWZ} \JL, \DVN, X. Wang, and A. Zichichi,
\PRD{48}{93}{2062} and \PRD{52}{95}{142}.
\bibitem{LP}See \eg, P. Langacker and N. Polonsky, \PRD{50}{94}{2199}.
\end{thebibliography}
\end{document}